# Movie Box office Prediction via Joint Actor Representations and Social Media Sentiment

**Dezhou Shen**
Department of Computer Science
Tsinghua University
Beijing, CN 100084
*sdz15@mails.tsinghua.edu.cn*

**Abstract**

In recent years, driven by the Asian film industry, such as China and India, the global box office has maintained a steady growth trend. Previous studies have rarely used long-term, full-sample film data in analysis, lack of research on actors' social networks. Existing film box office prediction algorithms only use film meta-data, lack of using social network characteristics and the model is less interpretable. I propose a FC-GRU-CNN binary classification model in of box office prediction task, combining five characteristics, including the film meta-data, Sina Weibo text sentiment, actors' social network measurement, all pairs shortest path and actors' art contribution. Exploiting long-term memory ability of GRU layer in long sequences and the mapping ability of CNN layer in retrieving all pairs shortest path matrix features, proposed model is 14% higher in accuracy than the current best C-LSTM model.

## 1      Introduction

With the development of social networks, actors' influence on social networks is increasing, and the influence of actors' consumption is spread through social networks, and the use of long-cycle, full-sample social network data to study the relationships between all actors, directors and films is a challenge in the field of social media research.

It is easy to see that the relationship between the box office and the director, actor, scriptwriter and producer is often non-linear, that is, it is difficult to use linear functions to build box office prediction models.

By analyzing the social network characteristics of movie actors, sentiment of movies in Sina Weibo posts, all pairs shortest path, it is easy to get the priority of each characteristic by analysis of feature importance, existing prediction algorithm only makes use of the film meta-data, lacks the study on the social network characteristics of the actors, though traditional explanation of the film box office prediction algorithm is sufficing, however the prediction effect is not satisfactory.

This paper presents a model, named FC-GRU-CNN, for movie box office prediction, combining characteristics of film meta-data, actors' social network measurement, all nodes pair shortest path, social network text sentiment and actors' art characteristics, proposed model is 14% higher in accuracy than the current best C-LSTM model.

## 2      Recent Work

The film industry is a multi-billion-scale industry, with China's box office exceeding 60 billion yuan in 2018, reaching 60.9 billion yuan, and predicting the acceptance and adoption of new films among the public is a challenging topic.

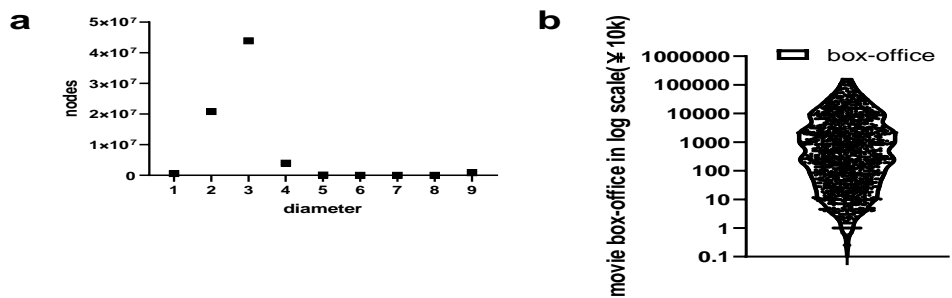

Figure 1    Actor shortest path feature distribution and movie box office distribution. (a) shortest path distance statistics (b) movie box office distribution for 1,296 films for the period 2011-2015.

## 2.1    Linear regression model based on social networks and movie meta-data

Zhu H[1] collected 13 features, such as search engine user data, social network fan number and movie meta-data of 20 films with more than 300 million box office in 2017, and constructed a partial least-squared model with a margin of error of 87.7% and an average absolute error of 26.6%.

Qiu X[2] used the 10 movie Sina Weibo texts and movie reviews of more than 1 billion at the box office in the Chinese film market in 2017 to calculate the web index and the film reviews to predict the box office.

## 2.2    Logical regression model based on online film review sentiment and movie meta-data

Some researchers use social network text sentiment to predict movie box office.

S Asur, et al.[3] collected 2.89 million texts related to 1.2 million users on Twitter in connection with 24 films, and concluded that the correlation coefficient $R^2$ was between 0.92 and 0.97 using the daily postings in the first seven days of release. The film box office forecast logic regression model with the distribution of social network posts was 0.97.

Jain V, et al.[4] collected 4800 Twitter texts of the movie name in the SNAP2009 (Stanford) dataset 2 weeks before release and 4 weeks after release. trained an sentiment al four-category model based on an 8-dollar grammar model with an accuracy rate of 64.4%. and eight films released in 2012 as test sets, using the sentiment al positive-negative ratio to predict the accuracy of the box office profit model by 50%.

Chi J[5] and others collected reviews from 150 films released in 2017 and found that the number of reviews was positively related to the box office. online reviews were not correlated with box office. and the number of reviews was significantly positive lying to the first week of the film's box office, and the relevance gradually diminished.

Josh M, et al.[6] collected the film review text of the film released by MetaCritic from 2005 to 2009, using film meta-data, ngram model, part-of-speech ngram model and dependency characteristics, to predict the first week box office, first week screen number and weekly revenue per screen, the determine of coefficient of the Elastic Net linear model is 0.671.

## 2.3    Support Vector Regression and Multi-Layer Neural Network Model of Fusion Review Sentiment s and Meta-data

From 2015 to 2017, Jiang L[7] collected 34 high-grossing films, using 10 features, such as film meta-data, director awards, and actor awards, and used voting feature selection algorithms to select features, training SVR (supporting vector regression) box office and scoring prediction models.

From 2012 to 2015, N Quader[8] collected 755 film reviews from four film review sites, including film ratings, MPAA categories, director's total box office, total box office, actor total box office, release time, budget, screen numbers, user reviews, reviews, and the sentiment al characteristics of review texts, training SVR, MLP(multi-layer neural network) model, with an accuracy of about 44.4%.

Table 1  Sina Weibo posts relating to movies.

| movie id | movie name | movie-related Sina Weibo text and retweet | movie participation actors list |
| --- | --- | --- | --- |
| 1 | Dafeng Zushi | The movie Big Peak Master: The trailer climaxes with wood? Suspense should be endless there are wood there? Want to catch a glimpse of wood? On September 5th, let's solve the ancient mirror mystery together with Xiao Li's catch- Zhang Wei, Sina Film, the trailer for the suspense version of "Dafeng Zushi" is directed by Zhang Wei, and the film "Dafeng Zushi" co-starring, such as Yu Ming, Xu Huishan, Han Xing An Zhenjing, releases the trailer for the suspense version. The main characters are all on the scene, and the killing caused by the ancient mirror mystery has become the biggest point of view, the complex relationship between the characters, the suspense of the plot is expected. The trailer for the suspenseful version of "The Great Peak Master" is a trailer for "Applause". Exposure Video: Trailer for the Suspense Edition of "Dafeng Master" | Wu Bin, Jiang Wei, Li Tongbo, Fan Chunli, Liao Weiwen, Sun Zhenguo, Gao Feifei, Ma Development, Li Zhenhe, Liang Runsheng, Li Min, Liu Ling, Teng Dejun, Zhang Zhi, Zhou Jian, Zhang Zhilan, Xue Gang, Liu Xiaotian, Peng Kai, Yu Ming, Lu Jing, Zheng Wei, Xu Huishan, Han Qi, Zheng Zhenlin, Yu Yuliang, Jiang Xingxuan, Liu Kailong, Lin Wenqing, Ma Yongzhou, Yu Shilin, Liang Xingjun, Zhuo Jinkui, Zhang Chenglong, Zhang Wei |

### 2.4  Deep Learning Text Classification

Chunting Zhou[9] and others proposed C-LSTM text classification model, using single-layer CNN to extract the high-order dimensional representation of text, and then connected the LSTM model to obtain sentence representation, the classification accuracy of the model is 87.8%

### 2.5  Summary

From the social network to predict the movie box office, feature selection includes the selection of user behavior characteristics, social network time characteristics, text sentiment al characteristics, classifiers more use of choice logic regression, support vector machine and neural network to predict the box office and comments.

## 3  Feature Extraction

### 3.1  Dataset

From 2011 to 2015, collected 3,669 films releasing Chinese mainland and 14,234 actors from the China-Box-Office, and collected the retweets and tweets of 8,508 actors on Sina Weibo social network.

The names of 7,369 actors were obtained from 14,234 actors participants records, as a dictionary of actors' names. There are 3,056 actors have opened account that can be collected from the social network.

A total of 1,298 films featuring social networking characters were extracted from 3,669 films, including at least one actor with social network ingress characteristics. Then entity alignment was performed, matching a total of 7,369 actors according to the actor's name entity alignment method, of which 3,056 had social network characteristics, while 4,313 actors did not have social network measurement data, which the network node represented as unknown node.

### 3.2  Feature Introduction

Movie Sina Weibo text features, extract the content of the Sina Weibo containing 1298 movie names from the social media and the posts text, calculate the sentiment al classification, as the sentiment characteristics of the Sina Weibo text.

Table 2　Sentiment characteristics of movie Sina Weibo posts

| movie id | movie name | release year | positive sentiment number | negative sentiment number | total sentiment | positive and negative sentiment ratio |
|---|---|---|---|---|---|---|
| 1 | Dafeng Zushi | 2014 | 135 | 33 | 168 | 4.0909 |

Actor social network measurement characteristics, measured the actor's social network as follows, including fan number, number of media, closeness, number of followers, number of blogs, average blog spacing, average number of tweets, number of retweets, movie mentions of Sina Weibo, total cash.

The shortest path feature of the social network structure of the actors, calculated the full-node shortest path of 8508 actors' social networks, and extracted the shortest path characteristics of 8508 dimensions of the 7,369 actors concerned. The calculation of the maximum minimum path for the full node is 8, for illustration purpose, the unreachable distance between nodes is recorded as 9, as Figure 1a.

Actors' art characteristics, the average box office as an indicator of the actor's box office impact, using the Chinese box office online actor's participation in the film box office data, to calculate the average film box office.

Using the box office data of the Chinese box office online actors participating in the film, the average film box office of the actorised, the average film box office as the actor's box office influence index, this characteristic as the actor's art characteristics. The average box office of 7,369 actors is counted from China's box office network as a feature of their art.

Sina Weibo posts text and movie actor participation name list refer to Table 1, for sentiment classification of texts and retweets, first calculate Bigram of the text, then calculate Tfidf of each Bigram, then a Logistic Regression classifier is trained. Use this model to classify sentiment of texts and retweets for each movie-related posts, then get positive number, negative number, total number and positive-negative ratio, the results are shown in Table 2.

In summary, the characteristics that need to be computed are movie meta-data, Sina Weibo sentiment, actors' participation name list. For actor representation, actors' network measurement, actors shortest path characteristics and art characteristics are needed. For movie partial representation, film meta-data, e. g., the year of release, and Sina Weibo sentiment characteristics, e. g., positive sentiment, negative sentiment, total sentiment and sentiment positive and negative ratio, are needed.

## 4　Feature Analysis

The following linear analysis of social network measurement features and the regression coefficient of film box office, and from the practical significance of the interaction between the characteristics and the film box office.

The positive characteristics associated with the box office in the linear regression algorithm are:
- Betweenness: the higher the number of actors, the higher the box office.
- Closeness: The higher the closeness of the actors, the higher the box office.
- Followers: The more other actors you pay attention to, the higher the box office.
- Number of posts: the more actors, the higher the box office.
- Average post interval: the greater the average hair interval, the higher the box office.
- Average characters of each posts: the longer the number of characters, the higher the box office, but the degree of relevance is not obvious.

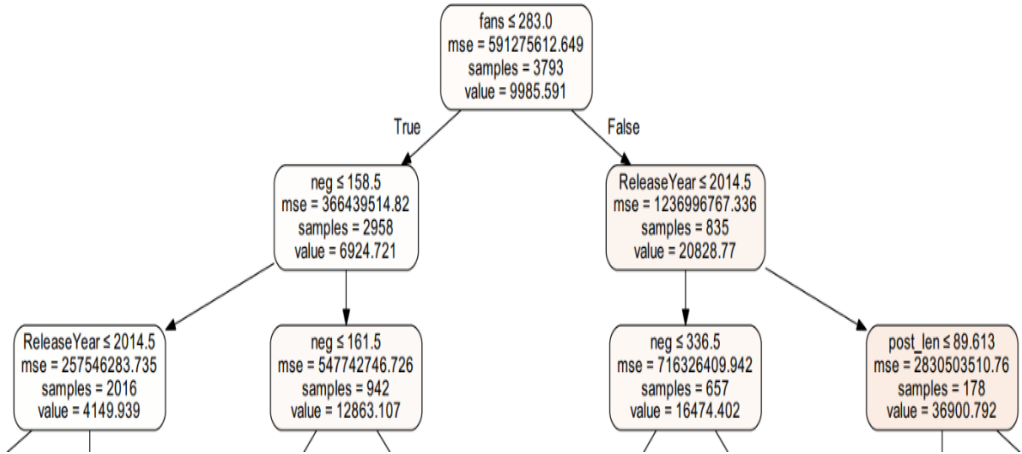

Figure 2   Actors social network meta-data decision tree visualization.

- Number of retweets: the more retweet posts, the higher the box office.
- Box office history: the higher the box office receipts of actors, the higher the box office attendance of participating in the film.

The negative correlation characteristics are:

- Number of fans: the more fans there are, the lower the box office is.
- Number of movie-related Sina Weibo posts: the larger number the actors' mentioned the movie in Sina Weibo, the lower the box office is.
- Co-occurrence: The more actors and other actors co-appears on Sina Weibo posts, the lower the box office is.

Then, calculated Pearson correlation of average box office per movie and actors social network measurement characteristics, the post and retweet correlation is -0.00108, social media Activity correlation is -0.03203, Co-occurrence correlation is 0.1508.

First of all, the actor's average box office per movie is clearly positively related to the influence of the actors, that is, the higher number of co-occurrence in Sina Weibo, the higher average box office per movie.

Secondly, the actors Sina Weibo activity and the average box office per movie showed a slight negative correlation, that is, the number of actors' followers, the number of tweets, the interval between the posts, the length of posts, the number of movie-related posts showed negative impact on average box office per movie.

Finally, the number of actors retweets and tweets is not related to average box office per movie, thus they are independent.

Then I use Sina Weibo text sentiment and actor social network measurement characteristics as features, then trained a decision tree to fit the movie box office, training set determination of coefficient is 0.9858, and the test set determination of coefficient is 0.9605. Actors social network meta-data decision tree visualization is illustrated as Figure 2.

**Definition 1 Gini Impurity**   A measure of how often randomly selected elements are incorrectly marked from a collection is a common CART (decision tree) algorithm evaluation indicator. If it is randomly marked based on the distribution of the label in a subset. Gini impurities can be divided by multiplying the probability of the item by the probability that it is misplaced. When in all conditions a node belong to a single category, it reaches its minimum value of zero.

In summary, the decision tree structure shows that the influence of the actors in the social network is the main factor affecting the box office results, followed by the movie-related Sina Weibo sentiment and the film meta-data is the secondary determinant, while the social network characteristics, movie-related Sina Weibo sentiment, film meta-data are the general determinants.

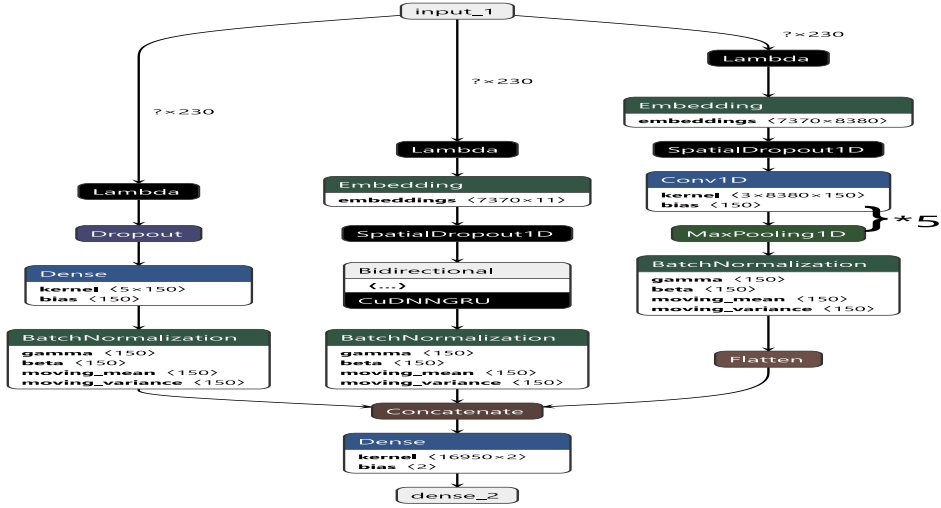

Figure 3 FC-GRU-CNN box office forecasting model structure

## 5  Actors Representation Learning with Social Network Measurement, Shortest path and Art Characteristics

**Definition 2 Network Embedding**   A network $G = (V, E)$, $V$ represents a set of nodes consisting of n nodes, and $E$ represents an edge set of m edges. For each edge $e \in E$, the ordered pair is $e = (u, v)$, $u, v \in V$ and the weight is $A_{ij}$. A network can be represented as an adjacent matrix, $\mathbf{A} = [A_{ij}] \in \mathbb{R}^{n \times n}$, and the goal of network embedding is to learn an all pairs representation, $\mathbf{U} \in \mathbb{R}^{n \times m}$, where $m (m < n)$ is an embedding dimension.

**Definition 3 Shortest Path Network Embedding**   In a network represented as a $G = (V, E)$, $V$ represents a node set of n nodes, and $E$ represents an edge set of m edges. For each edge $e \in E$, the ordered pair is $e = (u, v)$, $u, v \in V$ and the weight is $A_{ij}$. A network can be represented as an adjacent matrix, $\mathbf{A} = [A_{ij}] \in \mathbb{R}^{n \times n}$, and the goal of the shortest path network embedding is to learn an all pairs, shortest path representation, $\mathbf{U} \in \mathbb{R}^{n \times n}$, where n is an embedding dimension.

For the task of box office forecasting, because of the significant impact the actors on the box office, in addition to the measurement of the film meta-data and the sentiment of the movie social media, the actors' contribution is required to compute by leveraging the movie participants actors list. A film, led by several actors, obtained the names of 7,369 actors who participated in films between 2011 and 2015 are collected as an actor's dictionary, and then used the shortest path features of the actors' social network. Combining social network measurement features, the shortest path network embedding, and the art features to construct the actor representation vector. For actors who have no accounts on social networks, marked as unknown, and initialized with zero vectors. Actor representation vectors include social network measurement features of 10 dimensions, shortest path features of 8380 dimensions, art features of 1 dimension. In total, 10-dimensional measurement features include fans number, betweenness, closeness, followers, posts number, average posts interval, average word count per post, retweet number, movie-related post number, co-occurrence number, shortest path 8380 dimensions.

## 6  Machine Learning Classifier

Using the features extracted, the characteristics of participating in the calculation are film meta-data and movie sentiment: the year of release, the number of positive sentiment, the number of negative sentiment, the total number of Sina Weibo, the positive and negative ratio of sentiment, and the training of machine learning classifiers.

Table3    Random chosen 10 movies box office binary classification predictions in test set

| Movie name | Prediction | Actual | Box office(in 10K) |
| --- | --- | --- | --- |
| 盗剑 72 小时<br>72 hours of sword theft | A | A | 108 |
| 巴啦啦小魔仙<br>Ballera's Little Magic Fairy | B | B | 1916 |
| 缘来是咱俩<br>It's us. | A | A | 28 |
| 双生灵<br>Twin Spirits | B | B | 310 |
| 饭局也疯狂<br>The dinner is crazy, too. | B | B | 4381 |
| 地下室惊魂<br>Basement Horror | A | B | 704 |
| 铁塔油花浪漫曲<br>Tartar Oil Flower Romance | B | A | 1 |
| 破五<br>Breaking five | B | A | 1 |
| 绑架对门狗<br>Kidnapping the door dog | A | A | 1 |
| 北京爱情故事<br>Beijing Love Story | B | B | 40555 |

Machine learning methods such as Linear Regression, SVR, MLP, Logistic Regression, and Naive Bayes cannot converge on the data set, possibly because the sentiment on actors' social networks does not represent the commercial benefits of the film. Actors, as creators of films, have a role to play in the box office, but lack other important features that also have an impact on the box office, so machine learning fitting process can't finish in given time.

Using the CART decision tree regression algorithm to make box office prediction for the test set film, the experimental results show that the prediction model has a determination coefficient of 96.05% on the test set and 98.58% on the training set, which is 2.87% higher than the current best Sentiment-Network-Logistic model, and the average absolute deviation of the model is 0.1285 million, about 28.56% of the average box office.

The explanatories of the decision tree prediction model is more intuitive and more interpretable than linear regression, logical regression, deep neural network and other models, according to the prediction model structure schematic of the movie box office, the characteristics can be sorted according to the importance, so the social network structure characteristics are the most important, the social network text sentiment al characteristics, the film meta-data are important, and the social network characteristics are generally important.

With Qiu X's logistic regression model on sentiment and twitter index algorithm's 95.71% in determination of coefficient, the proposed method in this paper has a 2.87% improvement, 98.58%.

Table 4  FC-GRU-CNN model's performance

| Model | precision | recall | F1-value |
|---|---|---|---|
| C-LSTM[9] | 0.6154 | 0.6154 | 0.6154 |
| FC+GRU+CNN | 0.7500 | 0.7500 | 0.7500 |

## 7   A Deep Learning Classifier: FC-GRU-CNN Model

Look at the box office distribution of 1,296 films released between 2011 and 2015, as shown in Figure 1b.

Consider box office forecasts as a binary classification task. In order to make the data evenly distributed, the film data was pre-processed by using the median box office of 263.5(10K) as the dividing line. So for the movies with box office less than 2.635 million are classified as A-class, more than 2.635 million movies are classified as B-class.

In the data set division, the data was randomly selected using the 80-20 segmentation ratio, and the training set of 1036 films and the test set of 260 movies were obtained.

Using the actors' social network representation, the actor is quantified, that is, the fusion of social network measurement features, the shortest path characteristics and the actor's art characteristics, the actor feature matrix, for actors without social network measurement data, that is, the unknown characteristics, zero-value initialization. The model structure is shown in Figure 3:

The whole model is divided into three parts:
- The first part is FC full connection layer, its input feature is the film meta-data and movie-related Sina Weibo sentiment features, in total of 5 dimensions.
- The second part is the GRU layer, the input feature is the actor social network measurement characteristics and art characteristics, the actor table is a long list, the longest 225 actors, for less than 225 actors using the front-end padding, the GRU layer is to solve the long-term memory loss of the structure.
- The last part is the 5-layer CNN structure, input for the actors' social network shortest path feature, similarly, the actor table is a long list, input less than 225 dimensions is fulfilled by padding, CNN has a good ability to capture multi-dimensional features, and use of Max-pooling layer to solve the problem of multi-dimensional feature extraction.

Finally, the three parts are merged into one layer and then connect to a Softmax layer, and the $L_2$ regularization method is adopted to avoid overfitting.

In summary, I choose randomly 10 movies of in test set, as shown in Table 3. The accuracy of the binary classification model in training set reached 99.61%, the accuracy in the test set is 75%, compared with other prediction algorithms, see Table 4.

From the practical point of view, the model of FC-GRU-CNN proposed in this paper converges on the training set and achieves better prediction results in the test set, so it can provide decision-making advice on the task of selecting suitable actors for the film, and in the task of film box office prediction, 75% accuracy can be used for box office prediction tasks.

## 8   Conclusion

Based on the film box office prediction of actor's Sina Weibo sentiment and social network characteristics, this paper discussed the relationship between the movie meta-data, the movie Sina Weibo text sentiment, the actor's social network feature, the shortest path feature and the actor's art characteristics. Then proposed a FC-GRU-CNN box office binary prediction model, compared with the current best C-LSTM model, the accuracy improves 14%, in a dataset of 1,296 movies released on China mainland collected from 2011 to 2015, proposed model has an accuracy of 75% in test set.

**References**

[1] Zhu H, Tang Z. Film Box Office Forecasting Methods Based on Partial Least Squares Regression Model[C]//Proceedings of the 11th International Conference on Computer Modeling and Simulation. ACM, 2019:234-238.

[2] Qiu X, Tang T Y. Microblog Mood Predicts the Box Office Performance[C]//Proceedings of the 2018 Artificial Intelligence and Cloud Computing Conference. ACM, 2018: 129-133.

[3] Sitaram A, Bernardo A H. Predicting the Future with Social Media[C]. IEEE/WIC/ACM International Conference on Web Intelligence and Intelligent Agent Technology IEEE, 2010:492-499.

[4] Jain V. Prediction of movie success using sentiment analysis of tweets[J]. The International Journal of Soft Computing and Software Engineering, 2013, 3(3):308-313.

[5] Chi J, Gu E, Li K. How Does Online Word-of-Mouth Impact Movie Box Office: An Empirical Investigation of China[C]//Proceedings of the 2019 International Conference on Artificial Intelligence and Computer Science. ACM, 2019:730-735.

[6] Joshi M, Das D, Gimpel K, et al. Movie reviews and revenues: An experiment in text regression[C]//Human Language Technologies: The 2010 Annual Conference of the North American Chapter of the Association for Computational Linguistics. Association for Computational Linguistics, 2010:293-296.

[7] Jiang L, Wang Z. Predicting Box Office and Audience Rating of Chinese Films using Machine Learning[C]//Proceedings of the 2018 International Conference on Education Technology Management. ACM, 2018:58-62.

[8] Quader N, Gani M O, Chaki D, et al. A machine learning approach to predict movie box office success[C]//2017 20th International Conference of Computer and Information Technology (ICCIT). IEEE, 2017:1-7.

[9] Zhou C, Sun C, Liu Z, et al. A C-LSTM neural network for text classification[EB/OL]. arXiv preprint arXiv:1511.08630, 2015.

[10] Adhikari A, Ram A, Tang R, et al. Rethinking complex neural network architectures for document classification[C]//Proceedings of the 2019 Conference of the North American Chapter of the Association for Computational Linguistics: Human Language Technologies, Volume 1 (Long and Short Papers). 2019:4046-4051.

[11] Dai J, Jin L, Wang X. Factors Affecting the Box Office of Chinese Main-Melody Films Based on Big Data[C]//Proceedings of the 2019 International Conference on Artificial Intelligence and Computer Science. ACM, 2019:741-744.

[12] Gazley A, Clark G, Sinha A. Understanding preferences for motion pictures[J]. Journal of Business Research, 2011, 64(8):854-861.

[13] Jiang L, Wang Z. Predicting Box Office and Audience Rating of Chinese Films using Machine Learning[C]//Proceedings of the 2018 International Conference on Education Technology Management. 2018:58-62.

[14] Hennig-Thurau T, Houston M B, Walsh G. Determinants of motion picture box office and profitability: an interrelationship approach[J]. Review of Managerial Science, 2007, 1(1):65-92.

[15] Topf P. Examining Success in the Motion Picture Industry[J]. The Park Place Economist, 2010, 18(1):15.

[16] Gazley A, Clark G, Sinha A. Understanding preferences for motion pictures[J]. Journal of Business Research, 2011, 64(8):854-861.

[17] Pangarker N A, Smit E. The determinants of box office performance in the film industry revisited[J]. South African Journal of Business Management, 2013, 44(3):47-58.